# تخمین میزان تخلخل در سنگ‌های ساختمانی بر پایه الگوی دودویی محلی یک بعدی و روش نرمال‌سازی تصویر

شروان فکری ارشاد*

استادیار دانشکده مهندسی کامپیوتر، واحد نجف آباد، دانشگاه آزاد اسلامی، نجف آباد، ایران
مرکز تحقیقاتی مه داده، واحد نجف آباد، دانشگاه آزاد اسلامی، نجف آباد، ایران
پست الکترونیکی: fekriershad@pco.iaun.ac.ir

**چکیده**

تاکنـون روش‌هـای متنوعـی بـرای تشـخیص عیوب سـطحی بر پایه تحلیل بافت تصویر ارائه شـده است. یکی از روش‌هایـی کـه ویژگی‌هـای مناسـبی را بـرای تحلیل بافت تصویر ارائه می‌کند، الگوی دودویی محلی اسـت. بر اسـاس مفهوم عیوب سـطحی، می‌تـوان تخلخل[1] در سنگ را در زمره عیوب سـطحی برشـمرد. در این مقاله روشی برای آشکارسـازی و تخمین میزان تخلخل در سنگ‌های سـاختمانی، بر پایه الگوی دودویی محلی یک بعدی ارائه شـده اسـت. روش ارائه شـده مشـتمل بر دو مرحله است. در مرحلـه آمـوزش، ابتدا عملگر الگوهـای دودویی محلی یک بعدی بر روی تصویر فاقد تخلخل اعمال شده و بردار ویژگی مبنا استخراج می‌شود. سپس تصویر پنجره بندی شده و برای هر پنجره بردار ویژگی مجدداً جداگانه استخراج می‌گردد. با مقایسـه عدم شباهت بردارها با بردار مبنا بر اساس معیار نسبت درستنمایی لگاریتمی، حد آستانه سالم بودن بـه‌دسـت می‌آید. در مرحله آشکارسـازی، تصویر آزمون پنجره بنـدی شـده، بـردار ویژگی هر کدام استخراج شده و بر اساس حد آستانه فوق، پنجره‌های حاوی تخلخل شناسایی می‌شوند. در نهایت میزان تخلخل در الگوی عیب تولیدی، بر حسـب درصد محاسبه می‌شود. جهت افزایش نرخ کشـف، یک مرحله پیش‌پردازش جهت نرمال‌سـازی تصاویر بر پایه روش روپوش شـبکیه تک میزانی ارائه شده است. نرخ کشف بر روی سه نوع سنگ ساختمانی تراورتن کرم، تراورتن پرتغالی و تیشه‌ای به ترتیب ۹۷٫۳۳، ۹۸٫۰۶ و ۹۵٫۸۲ حاصل شـد. پیچیدگی محاسـباتی کم، توانایی برخط بودن و حساسیت کم به نوفه[2] از جمله دیگر مزایای روش ارائه شده به شمار می‌روند.

**واژه‌هـای کلیـدی:** الگوی دودویی محلی، تشخیص عیوب سطحی، تخلخل، روپوش شبکیه تک میزانی، پردازش تصویر

## ۱- مقدمه

به هرگونه خلل، فرج، منفذ و سـوراخ در سـطح سـنگ، تخلخل می‌گویند. در سـنگ‌های ساختمانی، میزان تخلخل

---

2- Noise

1- Porosity

* نویسنده مسئول

اهمیت بیشتری دارد زیرا می‌تواند در مرغوبیت و کیفیت سطح تمام شده بنا از دیدگاه معماری تاثیرگذار باشد. از طرفی دوام و بقای ساختمان پس از گذشت زمان نیز به شدت به میزان تخلخل وابسته است چرا که وجود تخلخل در سنگ‌های استفاده شده در نمای ساختمان ضریب آسیب‌پذیری آن را در برابر یخ‌زدگی و باران‌های اسیدی و حوادثی همچون زلزله تغییر می‌دهد. بنابراین در کارخانجات سنگبری، یکی از عوامل اصلی در درجه‌بندی کیفی سنگ‌های بریده شده، میزان تخلخل سطحی آن‌ها است. بر اساس مشاوره‌ای که در این مقاله از مهندسان عمران و ساختمان‌سازی گرفته شده است، میزان تخلخل سطحی سنگ به‌صورت نسبت مساحت سطح معیوب به مساحت کل سنگ بیان می‌گردد. این موضوع در معادله(۱)، نشان داده شده است.

$$\text{میزان تخلخل سطحی} = \frac{P.A(m^۲)}{S.A(m^۲)} \times ۱۰۰ \qquad (۱)$$

درمعادله (۱)، P.A مساحت سطح متخلخل و S.A مساحت کلی سنگ را نشان می‌دهند. هم اکنون در اکثر کارخانجات این عمل به‌صورت بصری و توسط کارگران باتجربه صورت می‌گیرد. امّا با توجه به طول زمان کار و سرعت جابجایی قطعات بریده شده بر روی غلطک‌های حمل، دقت کارگران به مرور زمان کاهش می‌یابد. بنابراین آشکارسازی و تخمین میزان تخلخل در سنگ‌های بریده شده به‌صورت خودکار می‌تواند در کاهش هزینه‌های مالی و زمانی و افزایش دقت موثر باشد. چند نمونه از تخلخل در سنگ‌های ساختمانی در شکل (۱) نشان داده شده است.

با توجه به مفهوم عیوب سطحی[3] و شکل ظاهری اکثر سنگ‌های ساختمانی، هرگونه خلل و فرج بر روی سنگ را می‌توان جزء عیوب سنگ دانست. بنابراین روشی که در این مقاله ارائه خواهد شد، در زمرهٔ روش‌های آشکارسازی عیوب[4] قرار می‌گیرد. تاکنون برای آشکارسازی و تخمین میزان تخلخل در سنگ‌های ساختمانی، روش‌های گوناگونی بر اساس فناوری‌های مختلف مطرح شده است که از آن جمله می‌توان به تشخیص تخلخل به کمک امواج ماوراء صوت توسط بویلانوآر و همکارانش در [۳۸] و کیفیت سنجی سنگ به کمک تزریق مواد شیمیایی توسط دورسی و همکاران در [۳۹]، اشاره کرد. ولیکن تحقیقات ما نشان داد که برای تشخیص تخلخل بر اساس روش‌های پردازش تصویر کار چندانی صورت نگرفته است. البته بخش متخلخل سنگ در واقع بخشی معیوب از سطح سنگ تلقی می‌شود و به همین دلیل می‌توان صورت مسئله این مقاله را در زمره روش‌های تشخیص عیوب سطحی اشیاء قرار دارد. تاکنون در زمینه آشکارسازی بخش‌های معیوب از بخش‌های سالم در تولیدات صنعتی کارهای زیادی صورت گرفته است که از آن جمله می‌توان به تحقیقاتی توسط هانزایی و همکاران در [۱] بر روی سرامیک، ناوارو و همکارانش بر روی چوب در [۲]، کوهلیپ و گراگ در [۳] بر روی چرم اشاره کرد. تحقیقات کارآمد دیگری همچون تحقیق غوشا و همکارانش در [۴] بر روی تشخیص عیوب بافتی بر روی انواع پارچه، سارگونار و سوکانس بر روی تولیدات مسطح فلزی در [۵]، تیوانا و همکارانش بر روی شیشه [۶] نیز در این همین حوزه قرار می‌گیرند.

به‌طور کلی در طراحی سیستم‌های آشکارساز خودکار عیوب که به آن‌ها سیستم‌های خودکار بازرسی مبتنی بر بینایی ماشین[5] هم گفته می‌شود، دو هدف اصلی مدنظر است:

الف-دقت روش ارائه شده در تشخیص محدوده وسیعی از عیوب

ب-حجم محاسباتی[6] روش که توانایی برخط بودن[7] سیستم به آن وابسته است. (در ادامه، این مقاله به تفصیل برای آشکارسازی تخلخل بررسی خواهد شد).

در راستای این اهداف، در زمینه آشکارسازی بخش‌های معیوب از بخش‌های سالم در تولیدات صنعتی، روش‌های گوناگونی ارائه شده است. صرف‌نظر از

---

3- Surface Defect
4- Defect Detection

5- Automatic Visual Inspection Systems
6- Computation Complexity
7- On-Line

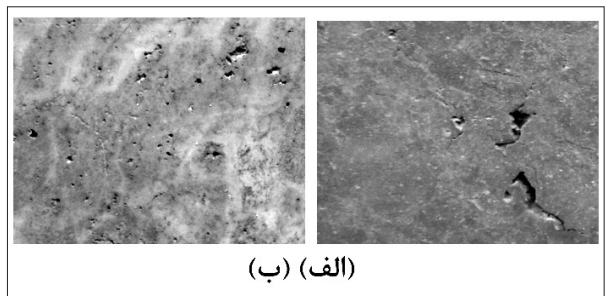

شکل۱: نمونه‌هایی از تخلخل
(الف) مرمر هرسین (ب) سنگ تراورتن پرتقالی

زمینه کاربردی هر یک از آن‌ها، می‌توان اکثر روش‌های مطرح شده را به ۴ گروه اصلی شامل روش‌های آماری، ساختاری، مبتنی بر مدل و مبتنی بر پایانه تقسیم کرد که در بخش دوم مروری بر روش‌های مطرح هر کدام از این حوزه‌ها خواهیم داشت. مولفه‌هایی مانند نرخ کشف، بار محاسباتی، نوفه و چرخش در تصویر از جمله چالش‌های اصلی در این حوزه است که اکثر روش‌های پیشین سعی کرده‌اند به نحوی آن‌ها را حل کنند. الگوهای دودویی محلی دو بعدی یکی از کارآمدترین روش‌ها برای تحلیل بافت یک تصویر و استخراج ویژگی‌هایی با قدرت تفکیک دهندگی بالا است. حساسیت به نوفه و بار محاسباتی از جمله ایرادات جدی است که می‌توان به این عملگر وارد کرد. در این مقاله ابتدا نسخه یک بعدی از الگوهای دودویی محلی در راستای حل این چالش‌ها ارائه شده و سپس از آنجا که روش پیشنهادی باید قابلیت آشکارسازی تخلخل در انواع سنگ‌های ساختمانی و همچنین توانایی برخط بودن را دارا باشد، بنابراین الگوریتمی با استفاده از ویژگی‌های الگوهای دودویی محلی یک بعدی ارائه می‌گردد. شایان ذکر است که برای صرفه جویی در زمان و کاهش هزینه‌ها، توانایی برخط بودن روش به شدت می‌تواند از جنبه کاربردی موثر باشد.

در بخش نتایج، تعداد مناسبی تصویر از سه نوع سنگ ساختمانی پر کاربرد شامل تراورتن کرم، تیشه‌ای و تراورتن پرتقالی تهیه شده است که شامل تصاویری با وجود تخلخل و عاری از هرگونه تخلخل طبق نظر کارشناسان مربوطه می‌باشند. سپس معیارهای نرخ کشف، نرخ حساسیت و نرخ ویژگی روش پیشنهادی بر روی این پایگاه تصاویر سنجیده شده است. در نهایت برای سه گروه سنگ ساختمانی به ترتیب دقت کشف ۹۷٫۳۳، ۹۸٫۰۶ و ۹۵٫۸۲ حاصل شد که نسبت به الگوهای دودویی محلی دو بعدی در تمامی موارد رشد دقت کشف به‌طور متوسط بیش از دو درصد می‌باشد. بار محاسباتی پایین‌تر عملگر الگوهای دودویی محلی یک بعدی نسبت به روش دو بعدی، نرخ حساسیت پایین‌تر و عدم حساسیت به چرخش و تغییر مقیاس تصویر از جمله مزایای روش پیشنهادی است که در بخش نتایج بحث خواهد شد.

### ۱-۱- ساختار مقاله

در ادامه این مقاله، ابتدا در بخش دوم کارهای مرتبط گذشته مرور شده و سپس روش الگوی دودویی محلی ابتدایی معرفی می‌گردد. سپس در بخش سوم، شکل بهبود یافته الگوی دودویی محلی دو بعدی نیز مطرح می‌شود. در بخش چهارم الگوی دودویی محلی یک بعدی ارائه می‌شود و در بخش پنجم، روشی برای استخراج بردار ویژگی معرفی می‌گردد. بخش ششم شامل روش ارائه شده برای تشخیص و آشکارسازی تخلخل در سنگ‌های ساختمانی است. در بخش هفتم، روشی برای نرمال‌سازی تصاویر ارائه شده است. نتایج تجربی پیاده سازی نیز در بخش هشتم آورده شده‌اند. در بخش نهم، روشی برای تخمین میزان تخلخل و درجه‌بندی کیفی سنگ‌های ساختمانی عنوان شده و در نهایت در دهمین بخش به بحث و نتیجه‌گیری پرداخته شده است.

### ۲- مروری بر کارهای گذشته

در راستای این اهداف، در زمینه آشکارسازی بخش‌های معیوب از بخش‌های سالم در تولیدات صنعتی، روش‌های گوناگونی ارائه شده است. صرف نظر از زمینه کاربردی هر یک از آن‌ها، می‌توان اکثر روش‌های مطرح شده را به ۴ گروه اصلی تقسیم کرد:

گروه اول: روش‌های آماری[8]
گروه دوم: روش‌های ساختاری[9]
گروه سوم: روش‌های مبتنی بر پالایه[10]
گروه چهارم: روش‌های مبتنی بر مدل[11]

روش‌هایی مانند تحقیق کانرز و همکاران برپایه ماتریس‌های هم‌رخ‌دادی[12] [8] و بویه و همکاران بر اساس تابع همبستگی[13] [9] همگی در گروه اول قرار می‌گیرند. به‌طور مثال بویه و همکارانش در تحقیق [9]، برای تشخیص عیوب بافتی در پارچه، روشی برحسب تابع همبستگی ارائه شده است. در این روش، در مرحله آموزش با انتخاب یک پنجره از تصویر سالم و حرکت دادن آن بر روی تصویر، حد آستانهٔ مناسبی به‌دست می‌آید و سپس در مرحله آزمایش به کمک حد آستانهٔ محاسبه شده، پنجره‌های معیوب آشکار می‌شوند. روش‌هایی که بر پایه ویژگی‌های لبه[14] [10]، نمایش اسکلت[15] [11] و یا اپراتورهای مورفولوژی [12] کار می‌کنند را می‌توان در گروه دوم قرار داد. به‌طور مثال ون و ژیا در [10]، برای تشخیص عیوب در تولیدات چرمی از ویژگی‌های استخراج شده از تصویر لبه‌یابی، استفاده کرده‌اند. در گروه سوم برای آشکارسازی عیوب از پالایه‌هایی از پیش طراحی شده، استفاده می‌شود. پالایه‌ها عموماً در دامنهٔ فرکانس[16]، دامنهٔ فضایی[17] و یا ادغامی از هر دو طراحی می‌شوند. کومار و همکارانش در [13] روشی برای آشکارسازی نقایص در پارچه‌های طرح‌دار با استفاده از پالایه‌های گابور ارائه کرده است. دراین روش ابتدا تصویر مورد بررسی از یک بانک پالایه عبور کرده و سپس خروجی پالایه آستانه‌گذاری می‌شود. در نهایت با ترکیب خروجی پالایه‌ها می‌توان به الگوی مناسبی از عیوب دست یافت.

در همین زمینه لینز و همکارانش[14] توانایی انواع پالایه‌ها را برای تشخیص عیوب بافتی مقایسه کرده‌اند که برای آگاهی بیشتر می‌توان به آن مراجعه کرد. گروه چهارم از روش‌ها را روش‌های مبتنی بر مدل می‌گویند. روش‌های مبتنی بر مدل‌های فرکتال توسط بوآ و همکاران [15]، مدل‌های میدان تصادفی[18] توسط کوهن و همکاران [16] و مدل‌های پسرفت خودکار[19] توسط مائو و جین [17]، همگی در این گروه قرار می‌گیرند. نتایج به‌دست آمده نشان می‌دهد که این گروه از روش‌ها عموماً در تحلیل بافت‌های طبیعی نسبت به تولیدات صنعتی بهتر عمل می‌کنند. لازم به توضیح است که دسته‌بندی فوق توسط ژی در [18] مطرح شده است.

## 2-1- الگوی دودویی محلی ابتدایی

برای تحلیل بافت تصویر، عملگرهای متنوعی تاکنون مطرح شده‌اند که هر کدام مزایا و معایب خاص خودشان را در کاربردهای گوناگون دارند. نتایج نشان داده است که در این میان، عملگر الگوهای دودویی محلی و نسخه‌های بهبود یافتهٔ آن، توانایی استخراج ویژگی‌هایی با قدرت تفکیک دهندگی بالایی را فراهم می‌کنند و نسبت به بسیاری از عملگرهای این حوزه بار محاسباتی کمتری نیز دارند. همچنین عدم حساسیت به چرخش تصویر از جمله دیگر مزایای عمده این عملگر است[7, 24]. در این مقاله، تلاش شده است تا ضمن حفظ ویژگی‌ها و مزایای این عملگر، نسخه‌ای بهبود یافته جهت تشخیص تخلخل در سنگ‌های ساختمانی ارائه شود که تا جای ممکن معایب و نقاط ضعف آن را هم پوشش دهد.

الگوی دودویی محلی[20] برای اولین بار توسط اوجالا و همکارانش در [21]، جهت تحلیل و طبقه‌بندی بافت تصویر ارائه شد. الگوی دودویی محلی عملگری غیرپارامتریک است که تباین محلی[21] و ساختار فضایی محلی[22] تصویر

---

8- Statistical
9- Structural
10- Filter Based
11- Model Based
12- Co-Occurrence Matrix
13- Correlation Function
14- Edge Features
15- Skeleton Representation
16- Frequency Domain
17- Spatial Domain

18- Markov Random Field
19- Autoregressive
20- Local Binary Patterns
21- Local Contrast
22- Local Spatial Structure

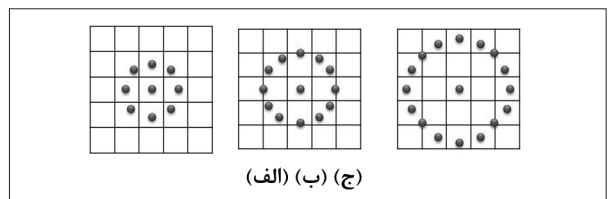

شکل ۲: نمونه‌هایی از همسایگی دایروی
(الف). ۸ =P و ۱ =R (ب). ۱۲ =P و ۱٫۵ =R (ج). ۱۶ =P و ۲ =R

را معرفی می‌کند. در این روش، ابتدا برای هر کدام از پیکسل‌های تصویر، یک همسایگی در نظر گرفته می‌شود و سپس برای محاسبهٔ میزان الگوی دودویی محلی در یک همسایگی از تصویر، از معادله (۲) استفاده می‌شود.

$$LBP_{P,R} = \sum_{i=0}^{P-1} S(g_i - g_c) 2^i \quad S(x) = \begin{cases} 1 & x \geq 0 \\ 0 & x < 0 \end{cases} \quad (۲)$$

در معادله (۲)، P نشان‌دهنده تعداد نقاط همسایه و $g_i$ شدت روشنایی نقاط همسایگی را نشان می‌دهد. همچنین $g_c$ شدت روشنایی نقطه مرکزی است. به‌طور معمول برای آنکه عملگر، نسبت به چرخش حساس نباشد، همسایگی به‌صورت دایره‌ای در نظر گرفته می‌شود. چند نمونه از همسایگی دایره‌ای با شعاع (R) در شکل(۲) نشان داده شده است. همان‌طور که در شکل(۲) مشخص است، مختصات برخی از نقاط همسایگی دقیقاً روی مرکز پیکسل قرار نمی‌گیرند. این سری از نقاط به کمک درون یابی[23] محاسبه می‌شوند.

خروجی عملگر الگوی دودویی محلی، عددی دودویی با P بیت اطلاعات است. با این اوصاف و همان‌طور که در شکل (۳)، نشان داده شده، نحوه نمایه‌گذاری پیکسل‌های همسایه می‌تواند منجر به تغییر مقدار الگوی دودویی محلی گردد.

برای حل این معضل، با چرخش عدد دودویی به‌دست آمده و انتخاب کمینه مقادیر ممکن، می‌توان مقدار یکتایی را به هر کدام از الگوهای محلی اختصاص داد[۲۱]. این مطلب در معادله (۳) نشان داده شده است.

$$LBP_{P,R}^{ri} = \min\{ROR\{LBP_{P,R}, \alpha\} \mid \alpha = 0,1,\ldots,p-1\} \quad (۳)$$

در معادله (۳)، عدم حساسیت اپراتور به چرخش با

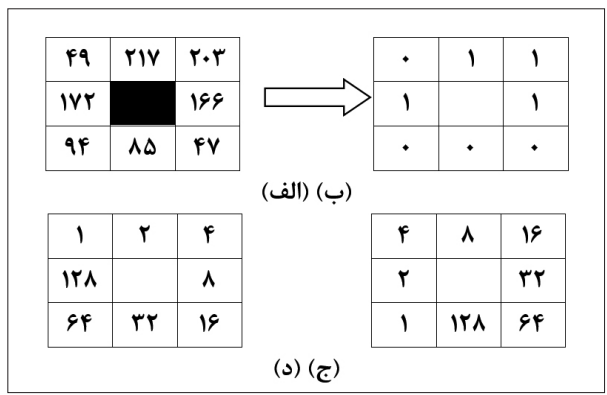

LBP ۱۴۲=۱۲۸+۸+۴+۲ LBP ۵۸=۳۲+۱۶+۸+۲

شکل۳: (الف) همسایگی مربعی ۳×۳ (ب) مقایسه پیکسل مرکزی با همسایه‌ها (ج) یکی از نحوه‌های نمایه‌گذاری و محاسبه LBP (د) یکی از نحوه‌های نمایه‌گذاری و محاسبه LBP

نماد ri نشان داده شده است. همچنین «ROR» نشان‌دهنده چرخش به سمت راست[24] است که «α» بار تکرار شده و حداقل اعداد به‌دست آمده به ازای «α»های بین صفر تا p-1 به‌عنوان الگوی دودویی محلی در نظر گرفته می‌شود.

## ۳- الگوی دودویی محلی دو بعدی

نتایج عملی اوجالا و همکارانش در [۲۲]، نشان داد که الگوی دودویی محلی که بدین طریق محاسبه می‌شود، با وجود توانایی مناسب برای تحلیل بافت تصویر، بار محاسباتی سنگینی را به سیستم وارد می‌کند و توانایی برخط بودن آن را کاهش می‌دهد. بنابراین اندک زمانی بعد، شکل بهبود یافته عملگر الگوی دودویی محلی توسط اوجالا و همکارانش [۲۳]، ارائه شد. در شکل بهبود یافتهٔ عملگر، معیاری به نام میزان یکنواختی[25] (همگنی) طبق معادله (۴) تعریف می‌شود.

$$U(LBP_{P,R}) = |s(g_{p-1} - g_c) - s(g_0 - g_c)| + \sum_{i=1}^{p-1}|s(g_i - g_c) - s(g_{i-1} - g_c)| \quad (۴)$$

همان‌گونه که در معادله (۴)، مشخص شده، میزان یکنواختی نشان‌دهنده تعداد جهش‌ها (جابجایی از صفر به یک و بالعکس) در شدت روشنایی نقاط موجود در همسایگی است. به‌طور مثال برای الگوی «۰۰۱۱۰۱۱»

---

23- Interpolation
24- Rotate Right
25- Uniformity Measure

میزان یکنواختی برابر با ٤ است. در این روش الگوهایی که میزان یکنواختی آن‌ها کمتر از حدآستانهٔ میزان یکنواختی($U_T$) باشد، الگوهای یکنواخت و الگوهایی که میزان یکنواختی آن‌ها بیش از $U_T$ باشد، به‌عنوان الگوی غیریکنواخت تعریف می‌شوند. درنهایت نیز با توجه به این تعریف، میزان الگوی دودویی محلی بهبود یافته طبق معادله(٥) محاسبه می‌شود.

$$LBP_{P,R}^{riu_T} = \begin{cases} \sum_{i=٠}^{P-١} s(g_i - g_c) \ if \ U(LBP_{P,R}) \leq U_T \\ P + ١ \ otherwise \end{cases} \quad (٥)$$

همان‌گونه که از معادله (٥)، برمی‌آید، در شکل بهبود یافته الگوی دودویی محلی، به همسایگی‌های یکنواخت، برچسب‌هایی[۳٦] از صفر تا P و به همسایگی‌های غیریکنواخت برچسب ١+P اختصاص داده می‌شود. با توجه به اینکه در شکل بهبود یافته الگوی دودویی محلی به تمام همسایگی‌های غیریکنواخت برچسب یکسان الصاق می‌شود، بنابراین برچسب‌های الصاق شده به الگوهای یکنواخت موجود در تصویر باید اکثر الگوهای موجود در تصویر را پوشش دهد و الگوهای غیریکنواخت تنها بخش ناچیزی از الگوها را شامل شود. نتایج عملی تاجری پور و همکارانش در [۲٤]، نشان داد که چنانچه حد آستانه میزان یکنواختی (UT) برابر با P/٤ در نظر گرفته شود، درصد کمی از الگوها (کمتر از یک درصد) دارای برچسب غیریکنواخت خواهند بود.

با توجه به انتخاب همسایگی دو بعدی، شکل بهبود یافته الگوی دودویی محلی عملگری از درجه دوم است. همین موضوع پیچیدگی محاسباتی سیستم را افزایش می‌دهد. بنابراین در بخش بعد، فرم یک بعدی الگوی دودویی محلی جهت کاهش هزینه محاسباتی و افزایش توان طبقه‌بندی بافت تصویر، توسط ما ارائه شده است.

## ٤- الگوی دودویی محلی یک بعدی

در الگوی دودویی محلی دو بعدی، انتخاب همسایگی

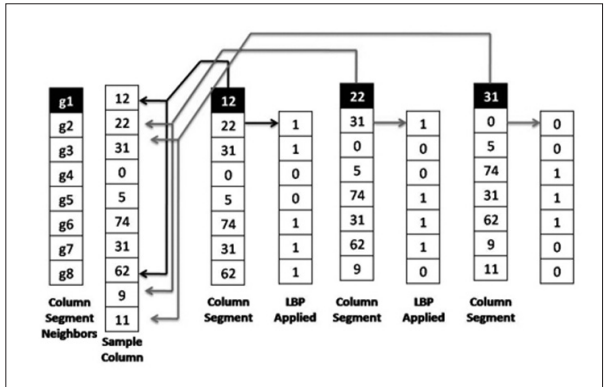

شکل ٤: اعمال عملگر الگوی دودویی محلی بر روی تکه ای عمودی از تصویر با قطعه ای به طول ٨

به‌صورت دایره‌ای، باعث عدم حساسیت اپراتور نسبت به چرخش تصویر می‌شود. ولیکن در سیستم‌های بازرسی، مسئله چرخش اهمیت بالایی ندارد، بنابراین نیازی به انتخاب همسایگی دایره‌ای نیست. از طرفی محاسبات درون یابی مورد نیاز در همسایگی‌های دایره‌ای بار محاسباتی بالایی را به سیستم وارد می‌کند که این امر می‌تواند توان برخط بودن سیستم را به شدت پایین بیاورد. بنابراین در این بخش نسخه جدیدی از الگوی دودویی محلی معرفی شده است که در آن همسایگی به‌صورت قطعه[۲۷] افقی (عمودی) در نظر گرفته می‌شود. با توجه به انتخاب همسایگی به‌صورت قطعات افقی (عمودی)، این نسخه الگوی دودویی محلی عملگری از درجه اول خواهد بود. بنابراین از این به بعد، این نسخه از الگوی دودویی محلی را الگوی یک بعدی می‌نامیم.

در الگوی دودویی محلی یک بعدی، سطح خاکستری اولین پیکسل از تصویر به ترتیب با سطوح خاکستری دیگر پیکسل‌های موجود در قطعه مقایسه می‌شود. مثالی از اعمال عملگر الگوی دودویی محلی یک بعدی بر روی یک تکه عمودی از تصویر در شکل (٤)، نشان داده شده است. در این روش میزان یکنواختی برابر با تعداد جهش‌ها (جابجایی از صفر به یک و بالعکس) در شدت روشنایی نقاط همسایه در قطعه افقی (عمودی) تعریف می‌شود. معادله (٦)، این مطلب را نشان می‌دهد.

---

27- Row(Column) Segment

26- Labels

$$U(LBP_L) = \left| s(g_L - g_1) - s(g_2 - g_1) \right| + \sum_{i=2}^{L-1} \left| s(g_i - g_1) - s(g_{i+1} - g_1) \right| \quad (6)$$

در معادله (۶)، L طول قطعه افقی(عمودی) را بر حسب پیکسل نشان می‌دهد و $g_1$ نشان‌دهنده سطح خاکستری اولین پیکسل در آن قطعه است. همچنین به دلیل تغییر شکل همسایگی، نماد $LBP_{P,R}$ به فرم $LBP_L$ تغییر می‌یابد. در این نسخه نیز الگوهایی که میزان یکنواختی آن‌ها کمتر از حدآستانهٔ میزان یکنواختی ($U_T$) باشد، الگوهای یکنواخت و الگوهایی که میزان یکنواختی آن‌ها بیش از $U_T$ باشد، به‌عنوان الگوی غیریکنواخت تعریف می‌شوند. بنابراین با توجه به این تعریف، میزان الگوی دودویی محلی یک بعدی برای قطعات افقی (عمودی) طبق معادله (۷) محاسبه می‌شود.

$$LBP_L^{U_T} = \begin{cases} \sum_{i=2}^{L} S(g_i - g_1) \; if \; U(LBP_L) \leq U_T \\ L \; Otherwise \end{cases} \quad (7)$$

همان‌گونه که از معادله (۷)، برمی‌آید، در الگوی دودویی محلی یک بعدی، به قطعات افقی(عمودی) یکنواخت، برچسب‌هایی از صفر تا L-۱ و به قطعات افقی (عمودی) غیریکنواخت برچسب L اختصاص داده می‌شود. برچسب‌های الصاق شده به الگوهای یکنواخت موجود در تصویر باید اکثر الگوهای موجود در تصویر را پوشش دهد و الگوهای غیریکنواخت تنها بخش ناچیزی از الگوها را شامل شود. نتایج عملی تاجری پور و همکارانش در [۲۴]، نشان داد که چنانچه حد آستانه میزان یکنواختی ($U_T$) برابر با L/٤ باشد، درصد کمی از الگوها دارای برچسب غیریکنواخت خواهند بود.

همان‌طور که اشاره شد، الگوی دودویی محلی دو بعدی به دلیل در نظر گرفتن همسایگی دو بعدی، عملگری از درجه دوم است ولیکن عملگر الگوی دودویی محلی یک بعدی به دلیل در نظر گرفتن همسایگی به‌صورت قطعات افقی و عمودی، عملگری از درجه اول است. به همین دلیل پیچیدگی محاسباتی آن نسبت به عملگر الگوی دودویی محلی دو بعدی بسیار کمتر است.

## ۵- استخراج بردار ویژگی

با توجه به توضیحات بخش چهارم، پس از اعمال عملگر الگوی دودویی یک بعدی بر روی تصویر، به قطعات افقی (عمودی) یکنواخت برچسب‌هایی از صفر تا L-۱ و به قطعات افقی (عمودی) غیریکنواخت برچسب L اختصاص داده می‌شود. برای استخراج بردار ویژگی، پس از اعمال عملگر بر روی تصویر و اختصاص برچسب‌ها به قطعات افقی(عمودی)، می‌توان احتمال برخورد به هر کدام از برچسب‌ها را به‌عنوان یکی از ابعاد بردار ویژگی محاسبه کرد. احتمال برخورد به هر برچسب به‌صورت نسبت تعداد قطعات با آن برچسب به تعداد کل قطعات تعریف می‌شود. این مطلب در معادله(۸) نشان داده شده است.

$$P_i = \frac{N_{P_i}}{N_{total}} \cdot \leq i \leq L \quad (8)$$

در معادله فوق، $P_i$ احتمال برخورد به برچسب i و $N_{pi}$ تعداد قطعات افقی(عمودی) با برچسب i در تصویر پس از اعمال $LBP_L$ بر روی کل تصویر را نشان می‌دهند. $N_{total}$ نیز تعداد کل قطعات افقی (عمودی) را نشان می‌دهد. بنابراین در مرحله استخراج بردار ویژگی می‌توان برای سطرها و ستون‌های تصویر دو بردار L+۱ بعدی استخراج کرد.

بدین ترتیب برای هر تصویر یک بردار ویژگی L+۱ بعدی برای قطعات افقی و یک بردار L+۱ بعدی هم برای قطعات عمودی قابل استخراج است. به‌عنوان مثال، در معادله (۹)، بردار ویژگی استخراجی برای قطعات افقی نشان داده شده است.

$$F_x = <P_0, P_1, \ldots, P_L> \quad (9)$$

## ۶- روش ارائه شده برای تشخیص تخلخل

همان‌طور که در مقدمه اشاره شد، به هرگونه خلل و فرج در سنگ، تخلخل گفته می‌شود. در این بخش، روش اصلی مقاله برای آشکارسازی تخلخل در سنگ‌های ساختمانی ارائه می‌گردد. نمودار شماتیک روش ارائه شده در این بخش در شکل (۵)، نشان داده شده است. همان‌طور که مشاهده می‌گردد، روش ارائه شده مشتمل بر

دو مرحله آموزش و آزمون (آشکارسازی) می‌باشد. در مرحله آموزش ابتدا یک تصویر فاقد تخلخل از سنگ مورد نظر دریافت شده (تصویر آموزش) و سپس بردار ویژگی برای کل آن تصویر بر اساس معادله (۹) استخراج می‌شود که آن را بردار مبنا می‌نامیم. سپس تصویر فاقد تخلخل به پنجره‌هایی با ابعاد مساوی و میزان همپوشانی دلخواه (در بخش نتایج بهترین ابعاد همپوشانی و پنجره بحث خواهد شد) تقسیم شده و بردار ویژگی برای هر پنجره به‌صورت مجزا استخراج می‌شود. در نهایت فاصله بردار ویژگی هر کدام از پنجره‌ها و بردار مبنا بر اساس نسبت درست‌نمایی لگاریتمی (در مقایسه با بسیاری از معیارهای فاصله و شباهت دقت بالاتری فراهم نمود) سنجیده شده و بیشینه فاصله به‌دست آمده به‌عنوان حد آستانه سالم بودن در نظر گرفته می‌شود. با توجه به تعریف عیب سطحی، تخلخل باعث می‌شود که بخش متخلخل تصویر، بافتی متفاوت با بافت بخش‌های سالم پیدا کند و چون الگوهای دودویی محلی تعریف کننده قوی از بافت تصویر هستند، این تفاوت در بردار ویژگی استخراجی توسط معادله (۹) خود را نشان خواهد داد. بنابراین در مرحله آزمون (آشکارسازی) تصویر مورد آزمایش پنجره‌بندی شده و بردار ویژگی برای هر پنجره استخراج می‌گردد. سپس میزان شباهت بردار ویژگی هر پنجره تصویر آزمایش با بردار مبنا سنجیده شده و چنانچه از حد آستانه سالم بودن بیشتر باشد، آن پنجره به‌عنوان یک پنجره حاوی عیب در نظر گرفته می‌شود. در ادامه جزییات روش ارائه شده بحث خواهد شد.

## ۶-۱- مرحله آموزش

همان‌طور که از نام این مرحله بر می‌آید، هدف اصلی در این مرحله استخراج بردار ویژگی و آموزش دادن سیستم است. در همین راستا، ابتدا تصویری از سنگ فاقد تخلخل تهیه می‌کنیم که در اصطلاح به آن تصویر سالم می‌گوییم. سپس عملگر الگوی دودویی محلی یک بعدی را بر روی کل تصویر سالم اعمال کرده و طبق معادله

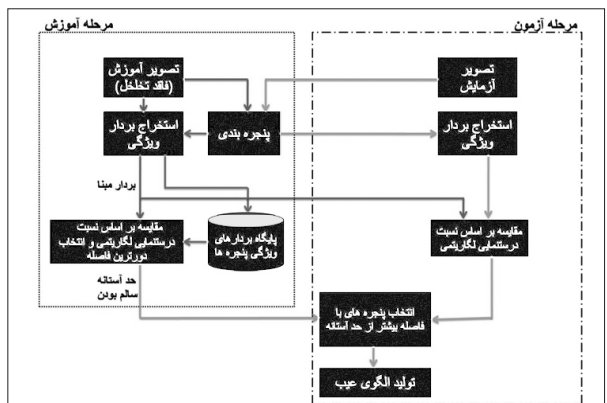

شکل ۵: تصویر شماتیک روش ارائه شده برای آشکارسازی بخش‌های متخلخل در سنگ ساختمانی

(۹)، دو بردار ویژگی برای آن استخراج می‌کنیم. بردارهای استخراجی را بردارهای مبنا می‌نامیم و به ترتیب بردار ویژگی استخراجی برای قطعات افقی را با نماد $M_x$ و بردار استخراجی برای قطعات عمودی را با $M_y$ نشان می‌دهیم. در ادامه تصویر سالم را به پنجره‌هایی با ابعاد W×W تقسیم کرده و برای هر کدام از پنجره‌ها به‌طور مجزا بردارهای ویژگی را استخراج می‌کنیم. اکنون می‌توان نسبت درست‌نمایی لگاریتمی بردار افقی (عمودی) هر کدام از پنجره‌ها را نسبت به بردار افقی (عمودی) مبنا، طبق معادلات (۱۰ و ۱۱)، محاسبه کرد.

$$L_{xk} = (S_{xk}, M_x) = \sum_{i=1}^{L+1} S_{iyk} \log\left(\frac{S_{ixk}}{M_{xi}}\right) \quad k = 1, 2, \ldots, N \quad (10)$$

$$L_{yk} = (S_{yk}, M_y) = \sum_{i=1}^{L+1} S_{ixk} \log\left(\frac{S_{iyk}}{M_{yi}}\right) \quad k = 1, 2, \ldots, N \quad (11)$$

در معادلات فوق، $S_{xk}$ و $S_{yk}$ به ترتیب بردارهای ویژگی افقی و عمودی استخراجی برای پنجره k ام است. همچنین N تعداد کل پنجره‌ها و i نشان‌دهنده بعد i ام از بردارهای ویژگی است. نسبت درست‌نمایی لگاریتمی معیار عدم شباهت است که همواره مثبت و حداقل مقدار آن برابر صفر خواهد بود و کمینه شدن آن نشان‌دهنده میزان شباهت با یک ردهٔ خاص می‌باشد. بنابراین بزرگ‌ترین مقدار محاسبه شده برای پنجره‌ها، به‌عنوان حد آستانه سالم بودن پنجره‌ها معرفی می‌شود. معادلات (۱۲ و ۱۳)، چگونگی محاسبه حدآستانه سالم بودن را به ترتیب برای قطعه بندی افقی و عمودی نشان می‌دهند.

$$T_x = \max(L_{xk}) \quad k = 1, 2, \dots, N \qquad (12)$$
$$T_y = \max(L_{yk}) \quad k = 1, 2, \dots, N \qquad (13)$$

### ۶-۲- مرحله آشکارسازی

اینک پس از محاسبه حد آستانه سالم بودن، عملاً سیستم بافت تصویر سالم را آموزش دیده است. بنابراین در این مرحله به آشکارسازی بخش‌های متخلخل در تصویر آزمایش(تصویری که ممکن است شامل بخش‌های متخلخل باشد) می‌پردازیم. در همین راستا، ابتدا تصویر ورودی را به پنجره‌هایی با ابعاد W×W تقسیم می‌کنیم. ابعاد پنجره‌ها می‌تواند بر روی دقت الگوریتم تاثیرگذار باشد. بنابراین پیش از تعیین ابعاد پنجره‌ها، در نظر گرفتن دو مورد اهمیت دارد:

الف) هر اندازه که ابعاد پنجره‌ها بزرگ‌تر در نظر گرفته شود، بردار ویژگی استخراج شده برای آن پنجره دارای مولفه‌های دقیق‌تری خواهد بود. ولیکن با افزایش ابعاد پنجره‌ها، نقش تخلخل‌های کوچک در ویژگی‌های استخراجی از پنجره کمرنگ‌تر شده و عملاً دقت الگوریتم در تشخیص تخلخل‌های کوچک کاهش می‌یابد.

ب) تعداد عملگرهایی که بر روی یک پنجره اعمال می‌شوند، به ابعاد پنجره و طول قطعات افقی (عمودی) وابستگی بالایی دارد. تاجری‌پور و همکارانش در [۲۴]، روشی برای تشخیص عیوب بافتی پارچه توسط الگوی دودویی محلی دو بعدی ارائه کرده‌اند. آن‌ها پیشنهاد داده‌اند که برای آن که بردار ویژگی استخراجی از هر پنجره، توانایی خوبی در معرفی ویژگی‌های آن پنجره داشته باشد، تعداد عملگرهای قابل اعمال از حدنصابی بالاتر باشد. بنابراین اگر ابعاد پنجره W×W و ابعاد قطعات افقی (عمودی) L باشند، تعداد عملگرهایی که بر روی یک پنجره اعمال می‌شود برابر با $W \times (W - L + 1)$ خواهد بود. اکنون اگر حد نصاب عملگرها را به‌طور مثال ۱۰۰ در نظر بگیریم، در نتیجه $(W^2 - W(L-1) - 100) > 0$ پس از پنجره‌بندی، برای هر کدام از پنجره‌ها به‌طور مجزا بردارهای ویژگی را استخراج می‌کنیم. سپس نسبت درستنمایی لگاریتمی بردار افقی (عمودی) هر کدام از پنجره‌ها را نسبت به بردار افقی (عمودی) مبنا، طبق معادلات (۱۴و ۱۵)، محاسبه می‌کنیم.

$$D_{xk} = (R_{xk}, M_x) = \sum_{i=1}^{L+1} R_{ixk} \log\left(\frac{R_{ixk}}{M_{xi}}\right) \quad k = 1, 2, \dots, N \qquad (14)$$

$$D_{yk} = (R_{yk}, M_y) = \sum_{i=1}^{L+1} R_{iyk} \log\left(\frac{R_{iyk}}{M_{yi}}\right) \quad k = 1, 2, \dots, N \qquad (15)$$

در معادلات (۱۴و ۱۵)، $R_{xk}$ و $R_{yk}$ به ترتیب بردارهای ویژگی افقی و عمودی استخراجی برای پنجره k ام از تصویر آزمایش هستند. همچنین $D_{xk}$ و $D_{yk}$ به ترتیب نسبت‌های محاسبه شده برای قطعات افقی و عمودی را نشان می‌دهند. بنابراین هر پنجره‌ای که نسبت درستنمایی لگاریتمی آن برای قطعات افقی یا عمودی بیش از حدآستانه سالم بودن باشد، به‌عنوان پنجره حاوی تخلخل معرفی می‌گردد. معادله (۱۶)، این مطلب را برای پنجره kام از تصویر آزمایش نشان می‌دهد.

$$K_{th} \; Window = \begin{cases} \text{متخلخل} & if \; D_{xk} > T_x \; or \; D_{yk} > T_y \\ \text{دیگر حالات بدون تخلخل} & \end{cases} \qquad (16)$$

خروجی الگوریتم آشکارسازی تخلخل، تصویر دودویی خواهد بود که الگوی عیب تولیدی نامیده می‌شود. در الگوی عیب تولید شده، نقاط سیاه، نشان‌دهنده بخش‌های بدون تخلخل و نقاط سفید نشان‌دهنده نقاط متخلخل می‌باشد.

## ۷- نرمال‌سازی تصویر توسط روپوش شبکیه تک میزانی[28]

در سال ۱۹۷۱، لاند و مک کان [۲۵] این ایده را معرفی کردند که هر تصویر، حاصل ضرب دو مولفه روشن‌سازی[29] L(x,y) و انعکاس[30] R(x,y) است. معادله(۱۷)، این موضوع را نشان می‌دهد. در واقع روشن‌سازی، ویژگی‌های هندسی[31] صحنه (موقعیت منبع نور و...) و انعکاس، اطلاعات مربوط به شی[32] را شامل می‌شوند.

$$I(x, y) = L(x, y).R(x, y) \qquad (17)$$

---
28- Single Scale Retinex
29- Illumination
30- Reflectance
31- Geometric Properties
32- Object

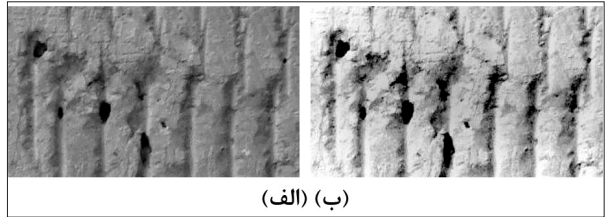

شکل ۶: (الف) تصویر اصلی (ب) پس از نرمال‌سازی

برای نرمال‌سازی تصویر یکی از اهداف اصلی کاهش میزان روشن‌سازی تصویر است که با استفاده از پالایهٔ بالاگذر ممکن است اما با توجه به این پیش فرض که تغییرات روشن‌سازی تصویر نسبت به موقعیت‌های[33] مختلف آرام و تغییرات انعکاس نسبت به موقعیت‌های مختلف سریع صورت می‌گیرد، نرمال‌سازی تصویر و حذف روشن‌سازی باید به گونه‌ای باشد که میزان انعکاس تصویر پس از پالایش به مقدار واقعی خود بسیار نزدیک باشد.

برای این موضوع لاند روشی به نام «روپوش شبکیه»[34] را ارائه داد. همان‌طور که از نام این روش برمی‌آید، روش روپوش شبکیه تلاش می‌کند که سیستم بینایی انسان را مدل کند. نکته اساسی در این روش این است که انعکاس $R(x,y)$ هر پیکسل یا تکه‌ای از تصویر برابر با خارج قسمت، تقسیم تصویر کلی $I(x,y)$ بر میزان روشن‌سازی محاسبه شده از همسایگان در آن تکه از تصویر است. روش «روپوش شبکیه تک میزانی» که توسط «جابسون» ارائه شد[26]، از یک کرنل گوسی[35] برای تخمین میزان روشن‌سازی همسایگان استفاده می‌کند. همچنین در این روش از مبدل لگاریتمی برای فشرده‌سازی بازه دینامیک تصویر استفاده شده است. معادله(۱۸)، این موضوع را نشان می‌دهد.

(۱۸)    $R_{SSR}(x, y) = \log I(x, y) - \log [F(x, y) * I(x, y)]$

در معادله، نماد (∗) نشان‌دهنده اپراتور پیچاندن[36] و $F(x,y)$ تابع گوسی است. همچنین $R_{SSR}$ تصویر نرمال شده می‌باشد. نمونه‌ای از نرمال‌سازی تصویر توسط روش روپوش شبکیه تک میزانی در شکل (۶)، نشان داده شده است.

از روش روپوش شبکیه تک میزانی در موارد کاربردی متنوعی همچون بهبود کیفیت تصویر توسط جابسون و همکاران[27]، تشخیص چهره توسط لسزینسکی و همکاران [28] استفاده شده است. بنابراین با توجه به توانایی بالای روش روپوش شبکیه تک میزانی برای نرمال‌سازی تصاویر و حذف روشن‌سازی، می‌توان یک مرحله پیش‌پردازش تصاویر برای روش ارائه شده آشکارسازی تخلخل در نظر گرفت. بدین صورت که ابتدا تمامی تصاویر آموزشی و آزمایشی توسط روش روپوش شبکیه تک میزانی نرمال شده و سپس وارد الگوریتم آشکارسازی گردند. توانایی مرحله نرمال‌سازی تصاویر در افزایش دقت آشکارسازی تخلخل در بخش نتایج نشان داده شده است. دلایل اصلی بهبود دقت تشخیص با استفاده از این روش نیز در بخش بعدی به تفصیل شرح داده شده است.

## ۷-۱- دلایل استفاده از روش روپوش شبکیه تک میزانی

در اکثر سیستم‌های بازرسی بصری، عموماً شرایط دامنه‌ای همچون جهت و شدت تابش نور، بازتاب نور، دستگاه عکس برداری وغیره، یکسان نیست. بنابراین تصاویری که برای مراحل آموزش و آشکارسازی گرفته می‌شوند، از لحاظ تباین[37]، روشن‌سازی و تابناکی[38]، متفاوت هستند. با توجه به این عوامل، پیش از مراحل آشکارسازی عیوب، نیاز به یک مرحله نرمال‌سازی تصاویر احساس می‌شود.

تاکنون روش‌های متنوعی برای نرمال‌سازی تصویر ارائه شده است که از آن جمله می‌توان روش تصحیح شدت روشنایی که توسط گونزالز و وودز در [29] مطرح شده یا روش برابرسازی هیستوگرام بر پایه بلوک توسط ژی و لام [30]، پالایه کردن همریختی توسط گونزالز و وودز [31] و برابر سازی انطباقی هیستوگرام توسط پیزر و آمبورن [32] اشاره کرد. نتایج تجربی ما نشان می‌دهد که به دلایل

---

33- Locations
34- Retinex( Retina+Cortex)
35- Gaussian Kernel
36- Convolution

37- Contrast
38- Brightness

زیر، روش روپوش شبکیه تک میزانی جهت نرمال‌سازی تصاویر در این کاربرد استفاده شود.

الف) روش روپوش شبکیه تک میزانی، روشن‌سازی در کل تصویر را به‌صورت یکسان حذف می‌کند. بنابراین در مرحلـه آموزش کـه عملگـر الگـوی دودویـی محلی بر روی کل تصویر سـالم، اعمال می‌شـود، چنانچه تصویر نرمال شود، بردارهای مبنا در هر بعد مولفه‌های دقیق‌تری خواهند داشت.

ب) روش روپوش شبکیه تک میزانی، روشن‌سازی را از تصویـر ورودی حذف کـرده و مولفه انعکاس تصویر ورودی را به‌عنـوان خروجـی برمی‌گرداند. همان‌طور که توضیـح داده شـد، مولفـه انعکاس، شـیء را در تصویر معرفی می‌کند. بنابراین قسمت‌های متخلخل نیز در همین بخش قرار می‌گیرند. با توجه به این موضوع، نرمال‌سازی انعکاس تصویر، منجر به افزایش شباهت بردارهای ویژگی استخراجی از پنجره‌های متخلخل به یکدیگر شده و بالطبع دقت تشخیص را افزایش می‌دهد.

پ) همان‌طور که در بخش دوم اشاره شد، روش الگوی دودویی محلی، تباین محلی را معرفی می‌کند. بنابراین اگر میزان تباین در کل تصویر ثابت نباشـد، ممکن اسـت که بردارهای اسـتخراجی از پنجره‌های واقعاً مشابه، یکسان نباشـد. روش روپوش شبکیه تک میزانی، میزان تباین در کل تصویر را هم نرمال می‌کند.

## ۸- نتایج

در این مقاله روشـی برای تشـخیص و آشکارسازی تخلخل در سنگ‌های سـاختمانی ارائه شد. بنابراین برای بررسـی کیفیت روش پیشـنهادی، می بایست پایگاهی از تصاویر سـنگ‌های سـاختمانی تهیه گردد که به‌صورت عمومـی پایگاه رایجی در این حوزه قـرار ندارد. بنابراین ابتـدا ۶۰ تصویر از ۳ مدل سـنگ سـاختمانی به نام‌های تراورتن پرتغالی[39]، تیشـه‌ای[40] و تراورتـن کرم بی‌موج[41]

توسط یک دوربین دیجیتال با کیفیت ۵ مگاپیکسل تهیه شد. در مراحل تهیه تصاویر تلاش شد که شرایط تابش نور از جمله میزان درخشـندگی و جهت تابش ثابت باشد. در این مراحل یکی از متخصصان عمران نیز ما را همراهی کرد تا مشخص شود که تصاویر قطعا از سنگ‌های دارای تخلخل یا کاملاً سالم دریافت شود و احیاناً عیوب ظاهری دیگری همچون شکسـتگی، لک، ترک و غیره به اشـتباه به‌عنوان تخلخل انتخاب نگردد. روش ارائه شـده در این مقاله برای هر نوع خاصی از سـنگ‌های سـاختمانی به‌صورت مجزا حد آسـتانه سـالم بودن و بردار ویژگی مبنا را استخراج می‌کند، بنابراین تعداد انواع سـنگ‌های ساختمانی تاثیری بر کلیات روش و نحوه تهیه پایگاه تصاویر ندارد. بنابراین برای هر مدل سنگ جدیدی نیز به شرط تهیه تصاویر فاقد تخلخل امکان اجرای روش پیشنهادی این مقاله وجود دارد. همچنیـن به دلیل رعایت نکات فوق، تقریبـاً اکثر روش‌هایی که برای تشـخیص عیوب سـطحی از اطلاعـات بصری تصویر شامل رنگ، شکل یا بافت استفاده می‌کنند می‌توانند بر روی این پایگاه تصاویر اجرا گردند. شـایان ذکر است روش‌هایی که برای تحلیل از روش‌های تحلیل سـیگنال یا تصویربرداری در شرایط خارج از نور مرئی همچون ماوراء بنفش یا مادون قرمز استفاده می‌نمایند، قابلیت استفاده از این پایگاه تصاویر را نخواهند داشت.

یکی از معیارهایـی که عمومـاً برای محاسـبه کارایی روش‌های آشکارسـازی عیوب به کار می‌رود، نرخ کشف اسـت [۳۳] که در معادله (۱۹)، نشـان داده شـده اسـت. بنابراین پس از اعمال روش پیشنهادی بر روی تصاویر، نرخ کشف بـرای الگوهای عیب تولیدی محاسـبه شـد. میانگین نرخ کشف عیوب به تفکیک مدل‌های سنگ در سطر اول جدول(۱)، نشـان داده شده است. شایان ذکر است که ابعاد مختلفی برای پنجره بندی تصاویر آزمایش شد و مشـخص گردید که پنجره با ابعاد ۱۶×۱۶ دارای بالاترین دقت می‌باشد.

$$\text{نرخ کشف} = 100 \times \frac{N_{cc} + N_{dd}}{N_{total}} \qquad (19)$$

---
39- Orange Travertine
40- Hatchet
41- Wave-Less Cream Travertine

جدول ۱: میانگین نرخ کشف، نرخ حساسیت و نرخ ویژگی در ۳ مدل سنگ بر حسب درصد

| معیار \ نوع سنگ | تراورتن کرم | تیشه‌ای | تراورتن پرتقالی |
|---|---|---|---|
| نرخ آشکارسازی | ۹۵٫۶۰ | ۹۶٫۲۲ | ۹۵٫۷۴ |
| حساسیت | ۹۳٫۵۸ | ۹۴٫۳۷ | ۹۳٫۹۵ |
| خاص بودن | ۹۶٫۷۶ | ۹۶٫۷۱ | ۹۶٫۸۱ |

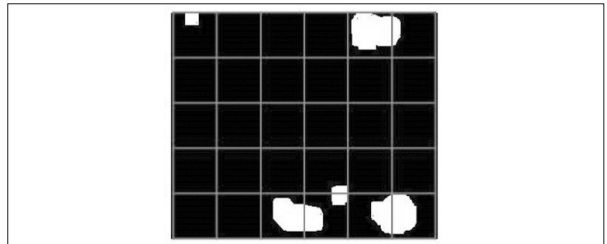

شکل ۷: نمونه‌ای از الگوی عیب تولید شده

جدول۲: میانگین نرخ آشکارسازی برای تصاویر پایگاه داده توسط عملگر الگوی الگوی دودویی محلی دو بعدی

| تعداد ویژگی‌ها | نوع عملگر \ نوع سنگ | تراورتن کرم | تیشه‌ای | تراورتن پرتقالی |
|---|---|---|---|---|
| ۱۰ | ۳و۸ | ۸۸٫۶۸ | ۹۱٫۶۴ | ۹۰٫۰۲ |
| ۱۸ | ۵و۱۶ | ۹۳٫۶۷ | ۹۲٫۲۷ | ۹۴٫۴۳ |
| ۲۶ | ۷و۲۴ | ۹۰٫۰۵ | ۹۱٫۳۲ | ۹۵٫۴۳ |
| ۱۸ + ۱۰ | ۳و۸ + ۵و۱۶ | ۹۳٫۲۲ | ۹۰٫۵۳ | ۹۱٫۱۴ |
| ۲۶ + ۱۰ | ۳و۸ + ۷و۲۴ | ۸۹٫۶۰ | ۸۹٫۰۳ | ۹۰٫۷۵ |
| ۲۶ + ۱۸ | ۵و۱۶ + ۷و۲۴ | ۸۹٫۴۴ | ۹۴٫۱۱ | ۹۲٫۴۶ |
| ۲۶ + ۱۸ + ۱۰ | ۳و۸ + ۵و۱۶ + ۷و۲۴ | ۸۵٫۵۴ | ۸۷٫۷۳ | ۹۱٫۳۷ |

در معادلـه (۱۹)، $N_{cc}$ به معنای تعـداد پنجره‌های واقعاً سالمی است که توسط روش نیز به‌عنوان پنجره سالم معرفی شـده‌اند و $N_{dd}$ تعداد پنجره‌هایی است که به‌عنوان پنجره متخلخل تشـخیص داده شـده و واقعاً نیز متخلخل بـوده انـد. در همیـن راسـتا، لازم به توضیح اسـت که بـرای محاسـبه نرخ آشکارسـازی، الگوی عیب تولیدی به پنجره‌هایی بدون همپوشانی با ابعاد ۱۶×۱۶ تقسیم گردید و پنجره‌ای که حداقل یک درصد (۳ پیکسـل) از آن معیوب بود به‌عنوان پنجره متخلخل در نظر گرفته شد. به‌طور مثال در شکل(۷)، اولین پنجره (سطر اول و ستون اول) به‌عنوان پنجره متخلخل در نظر گرفته می‌شود.

برای تمامـی الگوهای عیـب تولید شـده، میزان نرخ حساسـیت[۴۲] و نرخ ویژگی (خاص بودن[۴۳]) به ترتیب توسط معادلات (۲۰ و ۲۱)، محاسـبه شده و میانگین آن‌ها به ترتیب در سـطرهای دوم و سوم جدول (۱)، نشان داده شده است.

$$\text{حساسیت} = \frac{\text{مثبت واقعی}}{\text{منفی واقعی} + \text{مثبت واقعی}} \quad (۲۰)$$

$$\text{ویژگی} = \frac{\text{مثبت کاذب}}{\text{منفی کاذب} + \text{مثبت کاذب}} \quad (۲۱)$$

در معادلات فـوق، مثبت واقعی[۴۴] و مثبـت کاذب[۴۵] به ترتیب نشان‌دهنده تعداد پنجره‌های صحیح مثبت و صحیح منفی هستند. همچنین منفی کاذب[۴۶] و منفی واقعی[۴۷] هم به ترتیب تعداد پنجره‌های نادرست مثبت و نادرست منفی را نشان می‌دهند.

برای مقایسـه کیفیـت عملگر الگوی دودویی محلی یک بعدی، یکبار هم عملگر الگوی دودویی محلی دو بعدی بر روی تمامی تصاویر اعمال شـد و میزان دقت تشـخیص، حساسـیت و خاص بودن در تمامی الگوهای عیب تولیدی محاسـبه شـد. میانگین مقادیر محاسـبه شـده به تفکیک مدل‌های سـنگ به ترتیب در جداول(۲و۳و۴)، نشـان داده شده است. شـایان ذکر است که برای اعمال عملگر الگوی دودویی محلی دو بعدی، همسایگی به‌صورت مربعی و با ابعاد مختلف (۳×۳، ۵×۵، ۷×۷) در نظر گرفته شـدند. نتایج آشکارسازی در شکل(۸)، نشان داده شده است.

## ۸-۱- نتایج پس از نرمال‌سازی

در بخـش ۷، روش روپوش شـبکیه تک میزانی، برای نرمال‌سازی تصاویر ارائه شد. در این قسمت برای بررسی توانایی مرحله نرمال‌سازی، ابتـدا تمامی تصاویر پایگاه داده اعم از آموزش و آزمایش، نرمال شدند و سپس روش

---

42- Sensitivity
43- Specificit
44- True Positive
45- True Negative
46- False Positive
47- False Negative

جدول ۳: میانگین میزان حساسیت برای تصاویر پایگاه داده توسط عملگر الگوی دودویی محلی دو بعدی

| تعداد ویژگی‌ها | نوع سنگ / نوع عملگر | تراورتن کرم | تیشه‌ای | تراورتن پرتقالی |
|---|---|---|---|---|
| ۱۰ | ۳و۸ | ۸۵٫۹۸ | ۹۰٫۴۰ | ۸۹٫۴۵ |
| ۱۸ | ۵و۱۶ | ۹۲٫۲۴ | ۹۱٫۸۳ | ۹۳٫۳۱ |
| ۲۶ | ۷و۲۴ | ۸۹٫۳۲ | ۹۰٫۲۹ | ۹۳٫۲۶ |
| ۱۰ + ۱۸ | ۳و۸ + ۵و۱۶ | ۹۱٫۸۴ | ۸۷٫۷۲ | ۸۹٫۳۱ |
| ۱۰ + ۲۶ | ۳و۸ + ۷و۲۴ | ۸۷٫۶۰ | ۴۵٫۸۷ | ۸۹٫۶۶ |
| ۱۸ + ۲۶ | ۵و۱۶ + ۷و۲۴ | ۸۹٫۲۴ | ۹۲٫۹۱ | ۹۰٫۶۶ |
| ۱۰ + ۱۸ + ۲۶ | ۳و۸ + ۵و۱۶ + ۷و۲۴ | ۸۳٫۳۲ | ۸۶٫۴۰ | ۹۱٫۰۳ |

جدول۴: میانگین میزان خاص بودن تصاویر برای پایگاه داده توسط عملگر الگوی دودویی محلی دو بعدی

| تعداد ویژگی‌ها | نوع سنگ / نوع عملگر | تراورتن کرم | تیشه‌ای | تراورتن پرتقالی |
|---|---|---|---|---|
| ۱۰ | ۳و۸ | ۸۹٫۲۱ | ۹۲٫۲۴ | ۹۰٫۳۳ |
| ۱۸ | ۵و۱۶ | ۹۴٫۰۲ | ۹۲٫۷۲ | ۹۴٫۸۲ |
| ۲۶ | ۷و۲۴ | ۹۰٫۴۲ | ۹۲٫۸۱ | ۹۵٫۸۹ |
| ۱۰ + ۱۸ | ۳و۸ + ۵و۱۶ | ۹۳٫۶۰ | ۹۱٫۰۱ | ۹۱٫۷۷ |
| ۱۰ + ۲۶ | ۳و۸ + ۷و۲۴ | ۹۰٫۵۳ | ۹۰٫۷۷ | ۹۱٫۰۳ |
| ۱۸ + ۲۶ | ۵و۱۶ + ۷و۲۴ | ۹۰٫۷۵ | ۹۵٫۴۲ | ۹۲٫۹۲ |
| ۱۰ + ۱۸ + ۲۶ | ۳و۸ + ۵و۱۶ + ۷و۲۴ | ۸۷٫۴۰ | ۸۹٫۲۴ | ۹۲٫۱۱ |

پیشنهادی بر روی تصاویر نرمال شده، اعمال گشت. در نهایت نیز به کمک معادلات (۱۹و۲۰و۲۱)، میزان نرخ آشکارسازی، حساسیت و خاص بودن برای الگوهای عیب تولیدی محاسبه شد(جدول(۵). نتایج نشان می‌دهد که نرخ آشکارسازی با استفاده از یک مرحله پیش‌پردازش توسط روش روپوش شبکیه تک میزانی، افزایش چشم‌گیری می‌یابد. تعدادی از نتایج آشکارسازی در شکل(۸)، نشان داده شده‌اند.

## ۸-۲- مقایسه بار محاسباتی

در این مقاله از نسخه یک بعدی الگوهای دودویی محلی استفاده شده است. تفاوت نسخه یک بعدی و دو بعدی در نحوه تعریف همسایگی است. از دیدگاه تئوریک، در نسخه یک بعدی الگوهای دودویی محلی و تعریف همسایگان، می‌توان از دو حلقه مجزا از یکدیگر با مرتبه $O(n)$ استفاده کرد که n نشان‌دهنده تعداد همسایگان است. ولیکن در نسخه دو بعدی، همسایگی به‌صورت دایره‌ای تعریف می‌شود و برای پیاده‌سازی آن نیاز به استفاده از دو حلقه تو در تو برای بررسی مختصات افقی و عمودی همسایگان است که بار محاسباتی از مرتبه $O(n^2)$ می‌باشد. بنابراین از دیدگاه تئوریک بار محاسباتی نسخه یک بعدی نسبت به دو بعدی بسیار کمتر خواهد بود. از دیدگاه عملی نیز، آزمایشی در این مرحله اجرا کرده‌ایم. در این آزمایش۱۰ تصویر از سنگ‌های ساختمانی به‌صورت تصادفی انتخاب

شد و یکبار بردار ویژگی بر اساس نسخه یک‌بعدی و یکبار بر اساس نسخه دو بعدی برای هر تصویر استخراج شد. نتایج نشان داد که مدت زمان اجرا در تمام تصاویر بر اساس نسخه یک بعدی به‌صورت میانگین ۶۵٪ کمتر از نسخه دو بعدی بود.

## ۹- درجه‌بندی کیفی سنگ

برای درجه‌بندی کیفی سنگ‌های ساختمانی، میزان تخلخل در مقابل سایر شاخص‌هایی نظیر بافت، رنگ و برش از اهمیت بالایی برخوردار است. مهندسان ساختمان میزان تخلخل را، نسبت مجموع مساحت نقاط متخلخل بر کل مجموعه مساحت سنگ بیان می‌کنند. روشی که در این مقاله ارائه شد، به خوبی توانایی آشکارسازی نقاط متخلخل را دارد. همان‌طور که اشاره شد، هدف نهایی این مقاله سنجش میزان تخلخل سنگ و درجه‌بندی کیفی آن می‌باشد. بنابراین پس از اعمال روش ارائه شده بر تصویر سنگ و به‌دست آوردن تصویر دودویی الگوی عیب، می‌توان میزان تخلخل واقعی در سنگ را بر حسب درصد محاسبه نمود.

به این منظور می‌توان از معادلهٔ (۲۲) استفاده نمود:

$$\text{میزان تخلخل} = 100 \times \frac{\text{تعداد پیکسل‌های معیوب}}{\text{مجموع کل پیکسل‌ها}} \tag{۲۲}$$

جدول ۵: میانگین میزان نرخ کشف، نرخ حساسیت و نرخ ویژگی پس از نرمال‌سازی

| نوع سنگ / معیار | تراورتن کرم | تیشه‌ای | تراورتن پرتقالی |
|---|---|---|---|
| نرخ آشکارسازی | ۹۷٫۳۳ | ۹۸٫۰۶ | ۹۵٫۸۲ |
| حساسیت | ۹۵٫۲۱ | ۹۷٫۰۳ | ۹۴٫۶۵ |
| خاص بودن | ۹۸٫۶۶ | ۹۸٫۸۲ | ۹۷٫۲۳ |

جدول ۶: درجه‌بندی کیفی سنگ‌های ساختمانی بر حسب درصد تخلخل

| نوع سنگ / درجه کیفی | تراورتن کرم | تراورتن پرتقالی | تیشه‌ای |
|---|---|---|---|
| سوپر ممتاز | ۰٪ – ۵٪ | ۰٪ – ۳٪ | ۰٪ – ۷٪ |
| ممتاز | ۱۰٪ – ۵٪ | ۶٪ – ۳٪ | ۱۴٪ – ۷٪ |
| درجه یک | ۱۵٪ – ۱۰٪ | ۱۰٪ – ۶٪ | ۲۱٪ – ۱۴٪ |
| درجه دو | ۲۰٪ – ۱۵٪ | ۱۳٪ – ۱۰٪ | ۲۸٪ – ۲۱٪ |
| درجه سه | ۲۵٪ – ۲۰٪ | ۱۶٪ – ۱۳٪ | ۳۵٪ – ۲۸٪ |

همچنین بر اساس میزان تخلخل، می‌توان به‌صورت کاملا خودکار و برخط سنگ بریده شده را از لحاظ کیفیت درجه‌بندی نمود. عموماً مهندسان ساختمان سنگ‌های ساختمانی را از لحاظ کیفیت به ۵ دسته تقسیم می‌کنند. دسته اول که دارای کمترین میزان تخلخل و بالاترین کیفیت است، سنگ سوپر ممتاز نامگذاری می‌گردد. بقیۀ درجات به ترتیب ممتاز، درجه یک، درجه دو و درجه سه نامیده می‌شوند. نسبت میزان تخلخل و درجه کیفی سنگ، بسته به نوع سنگ و کاربرد آن متفاوت است. به‌طور مثال میزان تخلخل در سنگ تراورتن پرتقالی به دلیل کاربرد در بخش‌های تزئینی ساختمان، نسبت به سنگ تیشه‌ای، اهمیت بیشتری دارد که این موضوع از جدول(۶) به خوبی نتیجه می‌شود. در جدول (۶) این نسبت برای ۳ مدل سنگ ساختمانی نشان داده شده است. در جدول (۶) اعداد درون هر خانه نشان‌دهنده بازه‌ای از میزان تخلخل است که به کمک آن می‌توان درجه کیفی سنگ مورد نظر را با توجه به سطر مذکور طبقه‌بندی کرد. لازم به توضیح است که مقادیر ذکر شده در جدول (۶) با مشاوره یکی از کارشناسان این رشته (عمران)، به‌صورت تجربی تهیه شده است.

به‌طور مثال، میزان تخلخل در سنگ‌های ساختمانی نشان داده شده در شکل (۸)، به کمک معادله (۲۲) و بر اساس الگوی عیب تولیدی محاسبه شده و در جدول (۷) نشان داده شده است. همچنین درجه کیفی آن‌ها نیز با استفاده از جدول(۶)، تعیین شده است. شایان ذکر است که نتایج نشان داده شده در جدول (۷)، صرفا جنبه نمونه داشته و ارتباطی با دقت دسته‌بندی روش ارائه شده در این مقاله نمی‌باشد. در واقع نتایج این جدول نشان می‌دهد که روش ارائه شده امکان سنجش برخط میزان تخلخل (بر حسب درصد) و به تبع آن درجه‌بندی کیفی سنگ را فراهم می‌نماید.

با توجه به توضیحات بخش مقدمه، هم اکنون در اکثر کارخانجات، میزان تخلخل در سنگ‌های ساختمانی به‌صورت بصری و توسط افراد متخصص صورت می‌گیرد. ولیکن با استفاده از این روش، علاوه بر افزایش دقت تشخیص، هزینه‌های زمانی و مالی نیز کاهش یافته و میزان تخلخل به‌صورت عددی محاسبه می‌شود. برای بررسی کیفیت روش ارائه شده برای تخمین میزان تخلخل و درجه‌بندی کیفی سنگ، تمامی الگوهای عیب تولیدی در بخش نتایج یک بار توسط روش مذکور و یک بار هم توسط فرد متخصص، درجه‌بندی کیفی شدند. مقایسه نتایج نشان می‌دهد که در ۹۶٫۶ ٪ موارد (۵۸ تصویر از مجموع ۶۰ تصویر)، کیفیت سنگ مورد نظر به درستی درجه‌بندی شده است.

## ۱۰- بحث و نتیجه گیری

هدف اصلی این مقاله ارائه روشی مطمئن و سریع برای تخمین میزان تخلخل در سنگ‌های ساختمانی بود. در همین راستا، روشی مبتنی بر الگوی دودویی محلی یک بعدی و روپوش شبکیه تک میزانی ارائه شد. نتایج نشان داد که روش ارائه شده از دقت تشخیص بالایی برای اکثر مدل‌های سنگ‌های ساختمانی برخوردار است. از دیگر مزایای روش ارائه شده می‌توان به موارد زیر اشاره کرد:
الف) عملکرد الگوی دودویی محلی یک بعدی که در این

جدول ۷: تخمین تخلخل درالگوهای عیب تولید شده در شکل(۸)

| مدل سنگ \ نوع عملگر | تراورتن کرم | تراورتن پرتقالی | تیشه‌ای |
|---|---|---|---|
| الگوی دودویی محلی دو بعدی | ۲.۸۴ % (سوپر ممتاز) | ۳.۷۶ % (ممتاز) | ۵.۴۵ % (سوپر ممتاز) |
| الگوی دودویی محلی یک بعدی | ۱.۲۵ % (سوپر ممتاز) | ۳.۴۷ % (ممتاز) | ۳.۴۵ % (سوپر ممتاز) |

مقاله ارائه شد، بار محاسباتی کمتری نسبت به عملگر الگوی دودویی محلی دو بعدی دارد.

ب) عدم حساسیت به نوفه‌های کم به دلیل استفاده از روش پنجره بندی و در نظر گرفتن روابط همسایگی پیکسل‌ها

ج) روش ارائه شده برای تشخیص تخلخل به دلیل استفاده از مرحله آموزش، یک روش عمومی برای طبقه‌بندی دو رده‌ای است. بنابراین برای تشخیص عیوب در موارد کاربردی دیگر نیز می‌تواند استفاده شود.

د) معرفی بردار ویژگی نوینی که می‌تواند در بسیاری از زمینه‌های دیگر پردازش تصویر، جهت تحلیل بافت تصاویر استفاده شود.

و) دقت تشخیص بالای روش ارائه شده با استفاده از عملگر الگوی دودویی محلی دو بعدی، نشان می‌دهد که روش ارائه شده برای تشخیص تخلخل با انواع عملگرها و بردارهای ویژگی سازگاری خوبی خواهد داشت.

ه) دقت بالا در تشخیص میزان تخلخل و توانایی برخط بودن روش که کاهش هزینه‌های مالی و زمانی را برای کارخانجات سنگبری به ارمغان می آورد، از ویژگی‌های منحصر به فرد روش ارائه شده است.

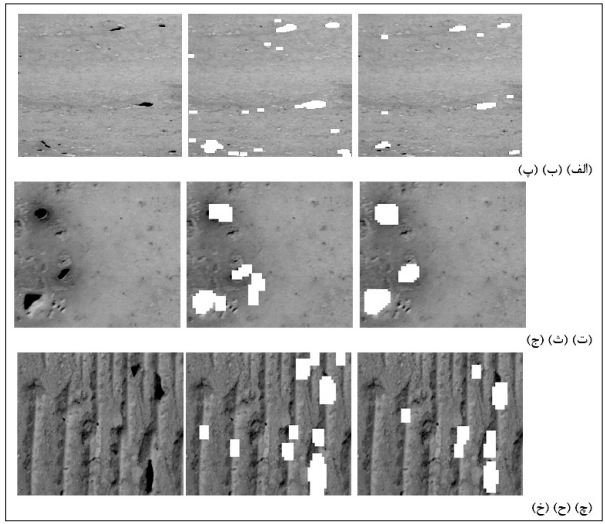

شکل ۸: (الف) تصویر اصلی از سنگ تراورتن کرم بدون موج (ب) الگوی عیب تولیدی با عملگر ۷×۷ دو بعدی (پ) الگوی عیب تولیدی با عملگر یک بعدی و طول قطعه ۸ پیکسل (ت) تصویر اصلی از سنگ تراورتن پرتقالی (ث) الگوی عیب تولیدی با عملگر ۳×۳ دو بعدی (ج) الگوی عیب تولیدی با عملگر یک بعدی و طول قطعه ۸ پیکسل (چ) تصویر اصلی از سنگ تیشه‌ای (ح) الگوی عیب تولیدی با عملگر ۵×۵ دو بعدی (خ) الگوی عیب تولیدی با عملگر یک بعدی و طول قطعه ۸ پیکسل

## منابع

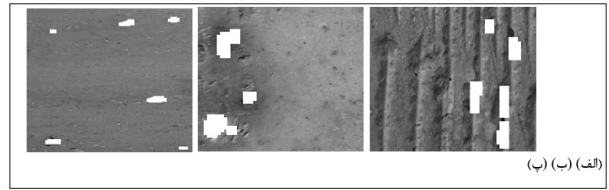

شکل ۹: برخی از نتایج آشکارسازی پس از نرمال‌سازی
(الف) الگوی عیب تولیدی برای شکل(۸. الف)
(ب) الگوی عیب تولیدی برای شکل(۸. د)
(پ) الگوی عیب تولیدی برای شکل(۸. ز)